\documentstyle[12pt,epsf]{article}
\begin{document}

\newcommand{\pa}{$\pi {-}a_1$}
\newcommand{\mch}{\stackrel{\circ}{m}}
\newcommand{\mpio}{\stackrel{\circ}{m}_\pi}
\newcommand{\Lch}{\left(\frac{\Lambda^2}{\Lambda^2+\mch^2}\right)^2}
\newcommand{\nh}{\stackrel{\circ}{h}_1}
\newcommand{\Io}{\stackrel{\circ}{I}_2}
\newcommand{\Do}{\stackrel{\circ}{\delta}}
\newcommand{\nhh}{\stackrel{\circ}{h}_2}
\renewcommand{\thefootnote}{\#\arabic{footnote}}

\thispagestyle{empty}

\noindent LPT 97-05 \hfill MAY 1997  

\noindent KFA-IKP(TH)-1997-10 

\noindent Coimbra University 260597

\vspace{1.5truecm}

\begin{center}
    {\large Quark-antiquark resonances in the NJL model\footnote{
        Supported in part by the
        grants PCERN /FIS/1034/95, PESO/S/PRO/1057/95,
        PRAXIS XXI/BCC/7301/96, and JNICT.}}
\end{center}

\vspace{0.6cm}

\begin{center}{V\'eronique Bernard\\
        {\small\em Laboratoire de Physique Th\'eorique, Universit\'e
         Louis Pasteur}\\
        {\small\em BP 28, F-67037 Strasbourg
        Cedex 2, France}\\

\vspace {0.4truecm}

        Alex H. Blin, Brigitte Hiller\\
        {\small\em Centro de F\'{\i}sica Te\'{o}rica, Departamento de
        F\'{\i}sica}\\
        {\small\em da Universidade de Coimbra, P-3000 Coimbra, Portugal}\\

\vspace {0.4truecm}

        Yuri P. Ivanov, Alexander A. Osipov\\
        {\small\em Joint Institute for Nuclear Research,
        Laboratory of Nuclear Problems}\\
        {\small\em 141980 Dubna, Moscow Region, Russia}\\

\vspace {0.4truecm}

        Ulf-G. Mei\ss ner\\
        {\small\em Forschungszentrum J\"ulich, Institut f\"ur 
         Kernphysik (Theorie)}\\
        {\small\em D-52425 J\"ulich, Germany}
       }
\end{center}


\vspace{1.5truecm}

\begin{abstract}
The problem of the $\rho$ meson resonance solution in the extended NJL model is
studied. It has been shown previously that the solution for the $\rho$ meson 
changes smoothly from a bound state to a ``virtual state" when the vector 
channel attraction becomes too weak to support bound state solutions. We have
found that together with these solutions a separate resonance branch appears 
which is not connected smoothly with the bound state solutions. At small values
of constituent quark mass, $m\sim 200$\ MeV, it corresponds to the physical 
vector meson resonance solution in the quark-continuum region. The $a_1$ and 
$\sigma$ resonances are also discussed.
\end{abstract}

\vspace{1truecm}

\newpage


\noindent {\bf 1.} 
We have become accustomed to think of resonance states in the extended 
Nambu--Jona-Lasinio (NJL) model \cite{Nambu:1961} as being solutions of the 
corresponding Bethe-Salpeter (BS) equations, analytically continued into the 
complex $p^2$-plane. This follows from our quantum mechanical experience in 
describing a system which can disintegrate. Such a system can have a very small
disintegration probability and is then known as a quasi-stationary or a 
resonance state. The propagator of a quasi-stationary state has always poles in
the complex $p_0$-plane which are situated close to the real axis. This method 
can also be applied to systems with a relatively large disintegration 
probability. Conceivably, we deal with such a kind of states in the case of the
strongly interacting quark-antiquark systems in the extended NJL model, if they
are located in the continuum region \cite{Takizawa:1990}. 

The question of the location of a complex pole which is close to the real axis 
can actually be resolved without going in the complex $p^2$ plane. In this 
letter we show how to do that. This simple procedure is quite general and can 
be used whenever one is interested in narrow resonance solutions of a known 
propagator or self-energy function of the corresponding state. The idea to find
out the resonance parameters without going into the complex $p^2$ plane has 
been used in the spectral-function (SF) method \cite{Weise:1990}-\cite{
Henley:1990}. It has been assumed that a maximum in the imaginary part of a 
meson propagator can be associated with a resonance mass. This approach has 
been criticized in \cite{Takizawa:1991}. It was argued there that although the 
SF method and the analytical continuation of the BS equations into the complex 
$p^2$ plane yield similar results for the $a_1$ meson, the same methods lead to
qualitatively different descriptions in the case of the $\rho$ meson. Using our
approach it is not difficult to understand why the SF method under certain 
circumstance may lead to wrong conclusions.

There are several reasons why the resonance solutions in a quark continuum
region are important. First, the $a_1$ meson always lies above the $q\bar{q}$ 
threshold. Second, the $\sigma$ meson is displaced in that region when the 
current quark mass is different from zero. Furthermore, the statement that 
there is no resonance solution for the $\rho$ meson \cite{Takizawa:1990, 
Takizawa:1991} requires  further investigation. In fact, it can not be excluded
that for solutions of the relativistic BS equation there exists a similar 
difference in the behaviour of $s$-- and higher partial waves as it is the case
in nonrelativistic potential theory. However, it is conceivable  that the 
relativistic BS equation exhibits some new features in comparison with a 
nonrelativistic theory. In this letter we will show that besides the bound and 
``virtual state" \cite{Takizawa:1991} solutions, which are smoothly connected 
at some value of the vector interaction strength $G_V$, there exists for all 
values of $G_V$ a resonance branch for the $\rho$-meson. We show that an 
overall fit to empirical data can be achieved either with large constituent 
quark masses ($\sim 400$\ MeV) \cite{Bernard:1988a} or small ones ($\sim 200$\ 
MeV). In the former case the physical $\rho$ meson appears as a bound $q \bar q
$ state. The resonance solution corresponds to the physical $\rho$ meson for 
small constituent quark masses. This finding lends credit to the approaches 
which use small values of non-strange quark masses $m$, i.e. when $m_\rho >2m$,
as it is frequently used in the literature \cite{Bijnens:1994}. It also shows 
that even so in the NJL model confinement is absent, one can still investigate 
some properties of states which a priori seem to lie outside the range of 
applicability of the model.

\vspace{1truecm}

\noindent {\bf 2.} Let the propagator of some state be defined by
\begin{equation}
\label{prop}
    \Delta (p^2) = f^{-1} (p^2) \,\, \, .
\end{equation}
It frequently happens that the function $f(p^2)$ becomes complex--valued for 
certain values of external momenta, and it is thus conceivable that the pole 
will be displaced into the complex $p_0$-plane. One assumes usually that near 
the pole the propagator behaves as
\begin{equation}
\label{pol}
    \Delta (p^2)\sim\frac{Z^{-1}}{p^2-M^2+i\gamma},\quad 
    Z=\left(\frac{\partial f(p^2)}{\partial p^2}\right)_{p^2=M^2-i\gamma} 
\end{equation}
where $\gamma$ is positive, so that for infinitesimal small $\gamma$ we make 
contact with the standard $i\epsilon$ prescription for the propagator poles. 
In this case the complex pole coordinates $p^2=M^2-i\gamma$ are interpreted as 
a mass, $M$, and a width, $\Gamma =\gamma /\omega (\mbox{\boldmath $p$})$, of a 
corresponding resonance state. Here $\omega (\mbox{\boldmath $p$})=\surd (
\mbox{\boldmath $p$}^2+M^2)$ is an energy. This interpretation follows from the
fact that the leading contribution of the Fourier--transformed two--point 
function at large time originates from the momenta associated with the 
propagator pole in the lower complex $p_0$-plane. 

If $\gamma$ is small ($\gamma\ll M^2$) the equation $f(M^2-i\gamma )=0$ can be
solved in the small width approximation, that is by expanding it around the 
point $M^2-i0$ in a Taylor series and restricting oneself to the terms linear 
in $\gamma$. This is possible if the function $f$ is analytic everywhere inside
a circle ${\cal C}_0$ with center at $M^2-i0$ and $M^2-i\gamma\in {\cal C}_0$. 
We assume that the branch cut of $f$ does not pass the circle ${\cal C}_0$.
In this case one obtains
\begin{equation}
\label{acs}
    \gamma =\frac{I_f}{R'_f}, \quad  R_fR'_f+I_fI'_f=0,
\end{equation}
where
\begin{eqnarray}
\label{not}
    R_f&=&\mbox{Re}f(p^2), \quad I_f=\mbox{Im}f(p^2), \nonumber \\
    R'_f&=&\frac{\partial\mbox{Re}f(p^2)}{\partial p^2}, \quad
    I'_f=\frac{\partial\mbox{Im}f(p^2)}{\partial p^2},
\end{eqnarray}
taken at $p^2=M^2$. Solving the second equation in (\ref{acs}) one can obtain 
the mass and the first one gives the width. It is, however, mandatory to check 
that the self-consistency requirement $\gamma\ll M^2$ is fulfilled.

\vspace{1truecm}

\noindent {\bf 3.} One can arrive at the equations (\ref{acs}) without going 
into the complex $p^2$ plane. Consider only real values of momenta. We look for
a point $q^2$ on the real $p^2$-axis such that in the neighbourhood of it the 
function $f(p^2)$ can be written in the form $f(p^2)\approx Z_0(p^2-q^2+i\gamma
)+{\cal O}(p^2-q^2)$. The function $f(p^2)$ can be expanded around some 
close--by but arbitrary point $q^2$,
\begin{equation}
\label{exp}
    f(p^2)=\mbox{Re}f(q^2)+i\mbox{Im}f(q^2)+(p^2-q^2)\left[
           \frac{\partial\mbox{Re}f(p^2)}{\partial p^2}+i
           \frac{\partial\mbox{Im}f(p^2)}{\partial
           p^2}\right]_{p^2=q^2}+\ldots
           \quad .
\end{equation}
The ellipses indicate terms of higher order in $(p^2-q^2)$. It follows that 
\begin{equation}
\label{asimp}
    f(p^2)=Z_0[p^2-q^2(1-\lambda )+i\gamma ]+\ldots
\end{equation}
where $\gamma$, $\lambda$ and $Z_0$ are given by
\begin{eqnarray}
\label{solv}
    \gamma (q^2)&=&\frac{I_fR'_f-R_fI'_f}{(R'_f)^2+(I'_f)^2}, \\
    \lambda (q^2)&=&\frac{R_fR'_f+I_fI'_f}{q^2[(R'_f)^2+(I'_f)^2]}, \\
    Z_0&=&\left(\frac{\partial f(p^2)}{\partial p^2}\right)_{p^2=q^2}
    \,\,\, . 
\end{eqnarray}
Thus, if $\lambda (q^2)$ is equal to zero one can identify the corresponding 
value  of $q^2$ with the mass of the resonance state and obtain the width of 
this state from Eq.(\ref{solv}). Indeed, if $\lambda (q^2)=0$ the equality 
$R_fR'_f+I_fI'_f=0$ is fulfilled. Inserting it in the expression for $\gamma$ 
and using that now $I_fR'_f-R_fI'_f=[(R'_{f})^{2}+(I'_{f})^{2}]I_f/R'_f$, we 
obtain a result fully coinciding with Eqs.(\ref{acs}). This shows that when the
points $q^2$ and $q^2-i\gamma$ are close to each other an approach with the 
self-consistency condition in the form 
\begin{equation}
\label{scc}
    \lambda (q^2) = 0    
\end{equation}
is equivalent to the method of the complex pole in the small width 
approximation. Actually, it is a simple consequence of the fact that the 
derivative of an analytic function $f(z)$ at the point $z_0$ does not depend on
the way which the point $z_0$ has been approached. In particular, it does not 
matter from which direction we calculate the value $f'(p^2)$ to obtain the 
result Eqs.(\ref{acs}), whether along the real axis or along the imaginary one.
The manner in which we have arrived at the formula Eq.(\ref{scc}) is not 
restricted by the requirement $\gamma\ll q^2$. It is clear, however, that at 
large values of $\gamma$, ($\gamma\sim q^2$), the relation of this solution 
with a complex pole location is not so direct. We conclude that in such type of
approach, the function $\lambda$ plays the crucial role to find a resonance 
state or to exclude it. 

In the case in which the function $f(p^2)$ is real, it is simple to see that 
$\lambda =R_f/(q^2R'_f)$ and the self-consistency requirement leads to $R_f=0$.
This is a well known fact that the mass of physical particles in quantum field 
theory is associated with the pole of the full propagator.

\vspace{1truecm}
 
\noindent {\bf 4.} Let us consider now the spectral-function method 
\cite{Weise:1990}-\cite{Henley:1990}. The spectral function is the imaginary 
part of the propagator $\Delta (p^2)$
\begin{equation}
\label{sf}
    \xi (p)=\mbox{Im}\Delta (p^2)=\frac{-I_f}{R_f^2+I_f^2} \quad .
\end{equation}
In this approach it is assumed that the parameters of the resonance can be
determined from the extremum of the function $\xi (p)$. The extremum condition 
is given by
\begin{equation}
\label{ext} 
    (R_f^2-I_f^2)I'_f-2R_fI_fR'_f = 0 \quad .
\end{equation}
By use of Eqs.(\ref{solv}) and (8) this condition can be written as  
\begin{equation}
\label{sfe}
    \lambda (q^2)+\frac{R_f(q^2)}{I_f(q^2)}\left(\frac{\gamma 
    (q^2)}{q^2}\right) = 0 \quad .
\end{equation}
It is obvious that this criterium is different from the requirement 
Eq.(\ref{scc}). Therefore, one must be fully aware that in the SF approach 
there is no direct relation with the complex pole method and from this point of
view can lead to incorrect results. This is in fact, very easily 
understandable. The difference between approaches 2 and 3 and the SF approach 
comes from the fact that in the SF approach one assumes that Z is real. Only in
that case is the Im $\Delta$, Eq.(\ref{pol}), maximum at $p^2=M^2$. Indeed if 
$I'_f=0$ then Eqs.(\ref{sfe}) and (10) are identical. The origin of a complex 
$Z$ factor comes from the appearance of a background term in the $T$ matrix so
that a difference between different approaches signals the presence of a 
non-negligible background. Whenever this happens it is useful to check the 
determination of the resonance parameters via the speed plot technique \cite{
Hoehler:1992}. 

\vspace{1truecm}
 
\noindent {\bf 5.} Consider now an extended NJL model with spin-1 mesons \cite{
Bernard:1994, Bernard:1996}\footnote{Other references can be found in \cite{
Bernard:1996}. We shall follow the notations of that paper here.}. For the
moment, let us postpone the rigorous solution of the mass equations and content
ourselves to discuss the simplest approach to the problem as outlined in 
section~3. We look for the resonance solutions for vector $\rho$, axial vector 
$a_1$ and scalar $\sigma$ mesons, on the basis of Eqs.(\ref{acs}). The 
corresponding self-energy functions can be written
\begin{eqnarray}
\label{frho}
    f_{\rho}(p^2)&=&\frac{3}{8G_V}-p^2J_2(p^2), \\
    f_{a_1}(p^2)&=&\frac{3}{8G_V}+6m^2I_2(p^2)-p^2J_2(p^2), \\
    f_{\sigma}(p^2)&=&\frac{\widehat{m}}{4mG_S}+(4m^2-p^2)I_2(p^2)
    \quad .
\end{eqnarray}
Here, $G_V$ and $G_S$ are the constants of the four--quark interaction of the 
vector, $V$, and the scalar, $S$, sectors. We also use the symbols $\widehat{m}
$ and $m$ for the current and constituent quark masses, respectively. The 
quark-loop integrals $I_2$ and $J_2$ are defined as
\begin{eqnarray}
\label{intij}
    &&I_2(p^2)=\frac{N_c}{16\pi^2}\int_0^1dy\int_0^{\infty}\frac{dz}{z}
               R(z)e^{\frac{i}{4}zp^2(1-y^2)}, \\
    &&J_2(p^2)=\frac{3N_c}{32\pi^2}\int_0^1dy(1-y^2)\int_0^{\infty}
               \frac{dz}{z}R(z)e^{\frac{i}{4}zp^2(1-y^2)}.
\end{eqnarray}
The Pauli-Villars regulator $R(z)$ \cite{Pauli:1949} is chosen in the form
\begin{equation}
\label{frz}
    R(z)=\left[1-(1+iz\Lambda^2)e^{-iz\Lambda^2}\right]e^{-im^2z}
    \quad .
\end{equation}
It corresponds to two covariant subtractions at large momenta $p^2=\Lambda^2
\gg m^2$ leading to a finite result for the $\rho$, $a_1$ and $\sigma$ 
self-energy diagrams. The first two of them contain the quadratically divergent
integrals. Using the Pauli-Villars method one can avoid these $\Lambda^2$ terms
from the self-energy diagrams. The following properties of the function $R(z)$ 
are important for that: $R(0)=R'(0)=0$. 

Evaluating the $I_2$ and $J_2$ integrals we get
\begin{equation}
\label{fi2}
    I_2(p^2)=\frac{3}{16\pi^2}\left\{\ln\left(\frac{\Lambda^2+m^2}{m^2}\right)
    +2\left[J(\bar\xi )-J(\xi )\right]+\left(\frac{\Lambda^2}{\Lambda^2+m^2}
    \right)\frac{J(\bar\xi )}{(\bar\xi -1)}\right\} \quad .
\end{equation}
\begin{eqnarray}
\label{fj2}
    J_2(p^2)&=&\frac{3}{16\pi^2}\left[\ln\left(\frac{\Lambda^2+m^2}{m^2}\right)
       +\left(2+\frac{1}{\bar\xi}\right)J(\bar\xi )-
       \left(2+\frac{1}{\xi}\right)J(\xi )+ \right. \nonumber \\
    &+&\left.\frac{1}{2\bar\xi}\left(\frac{\Lambda^2}{\Lambda^2+m^2}\right)
       \left(1+3\frac{J(\bar\xi )}{\bar\xi -1}\right)\right] \quad .  
\end{eqnarray}
Here 
\begin{equation}
\label{xi}
    \xi =\frac{p^2}{4m^2}, \quad \bar\xi =\frac{p^2}{4(\Lambda^2+m^2)}
    \,\,\,\, ,
\end{equation}
and the one-loop function $J(x)$ of the real variable $x$ is equal to
\begin{equation}
\label{fjx}
    J(x)=\left\{ \begin{array}{lll}
        \frac{1}{2}\sqrt{1-\frac{1}{x}}\ln\left(
        \frac{1+\sqrt{1-\frac{1}{x}}}{-1+\sqrt{1-\frac{1}{x}}}\right)
        & \mbox{if $x<0 \,\,\, ,$}\\   
        \sqrt{\frac{1}{x}-1}\arctan\frac{1}{\sqrt{\frac{1}{x}-1}}
        & \mbox{if $0<x<1 \,\,\, .$}
        \end{array}   
        \right.
\end{equation}
We remark that our function $J_2(p^2)$ coincides with the corresponding 
expression from \cite{Takizawa:1991}, precisely we have 
$16\pi^2p^2J_2(p^2)=9\Lambda^2 I_1(z)$, where $I_1(z)$ is given by Eq.(15) in 
\cite{Takizawa:1991}.

Let us fix the parameters of the model. There are four of them in all: $G_S, 
G_V, m$ and $\Lambda$. The constituent quark mass, $m$, is related to the 
current quark mass, $\widehat{m}$, by the gap equation of the model. We fix 
parameters of the model in order to obtain the pion decay constant, $f_\pi$, 
and the pion mass, $m_\pi$, close to their physical values, $f_\pi\simeq 93$\ 
MeV and $m_\pi\simeq 139$\ MeV, respectively. The other two input values are 
determined from the vector mesons sector such that one obtains values close to 
the empirical masses of $\rho$, $m_\rho \simeq 770\:\mbox{MeV}$ and $a_1$ 
meson, $m_{a_1}\simeq 1100\:\mbox{MeV}$. We investigate two cases, one with a 
large constituent quark mass ($m=390$\ MeV) leading to a bound state solution 
for the $\rho$ meson, and one with a small quark mass ($m=180$\ MeV), where the
physical $\rho$ meson appears as a resonant state in the $q\bar{q}$--continuum. 
We obtain for the large quark mass case: $G_S=9.41\:\mbox{GeV}^{-2},\ 
G_V=10.27\:\mbox{GeV}^{-2},\ \widehat{m}=3.9\:\mbox{MeV}$ and $\Lambda =1$\ 
GeV. With these parameters, we find $f_\pi =93.3\:\mbox{MeV},\ m_\pi =139\:
\mbox{MeV},\ m_\rho =776\:\mbox{MeV},\ m_{a_1} =1042-i358\:\mbox{MeV},
\ m_\sigma =439-i0.28\:\mbox{MeV}$. For the small quark mass case we get:
$G_S=0.86\:\mbox{GeV}^{-2},\ G_V=6.65\:\mbox{GeV}^{-2},\ \widehat{m}=1\:
\mbox{MeV}$ and $\Lambda =2.8$ GeV. With these parameters, we have $f_\pi 
=93.3\:\mbox{MeV},\ m_\pi =139\:\mbox{MeV},\ m_\rho =674-i300\:\mbox{MeV},
\ m_{a_1} =802-i289\:\mbox{MeV},\ m_\sigma =378-i2.6\:\mbox{MeV}$. 
Note that to determine these resonance properties, we have made use of 
Eqs.(\ref{acs}). 

\vspace{1truecm}

\noindent {\bf 6}. In order to search for resonances which lie in the complex 
plane on the second Riemann sheet, we introduce the following analytic 
continuation of the functions $J(\xi )$ and $J(\bar\xi )$. The function $J(\xi 
)$ has a branch point $P_0$ at $z=1$ and a cut along the real axis from $P_0$ 
to $+\infty$. Analytically continuing this function on the second Riemann sheet
we obtain the following analytic function ${\cal F}_0(z)$, where $z\in\mbox{
\boldmath $C$}$, 
\begin{equation}
\label{ff0}
    {\cal F}_0(z)=\frac{1}{2}\sqrt{1-\frac{1}{z}}\left[-i\pi +\log\left(
    \frac{1+\sqrt{1-\frac{1}{z}}}{1-\sqrt{1-\frac{1}{z}}}\right)
    \right] \,\, \, .
\end{equation}
We are looking for resonances whose masses are smaller than the cut-off
$\Lambda$ so that one does not cross the cut of the function $J(\bar\xi )$ 
which lies from $P_1$ to $+\infty$ where $P_1$ is at $\bar{z}=1$, that is 
at $z=1+(\Lambda^2/m^2)$. This function is thus given in the complex plane 
\mbox{\boldmath $C$} by the analytic function ${\cal F}_1(\bar{z})$ 
\begin{equation}
\label{ff1}
    {\cal F}_1(\bar{z})=\frac{1}{2}\sqrt{1-\frac{1}{\bar{z}}}\log\left(
    \frac{1+\sqrt{1-\frac{1}{\bar{z}}}}{-1+\sqrt{1-\frac{1}{\bar{z}}}}\right)
    \,\,\, ,
\end{equation}
where $\bar{z}\equiv zm^2/(\Lambda^2+m^2)$. Let us consider now the mass 
equations 
\begin{equation}
\label{cpe}
    f_{\alpha}(z)=0,\quad \alpha=\sigma , a_1, \rho \,\,\, .
\end{equation}
It is known\footnote{One can also prove this fact analytically.} that in the
case of $\sigma$ or $a_1$ mesons there exists only the resonance solution of 
these equations. To show them we shall plot functions $\omega_{\alpha}(G_V)$
and in some cases also $\eta_{\alpha}(G_V)$, where $(\omega_{\alpha}-i\eta_{
\alpha})^2=4m^2z_{\alpha}$ and $z_\alpha$ is a solution of the corresponding 
equation in Eq.(\ref{cpe}) at some value $G_V$. We shall call these functions a 
resonance branch of Eq.(\ref{cpe}) if they are a mapping of resonance solutions
of this equation at different values of $G_V$. The corresponding formulae are
\begin{equation}
\label{trs}
    \omega_\alpha =mK_\alpha ,\quad\eta_{\alpha}=-2mK_\alpha^{-1}\mbox{Im}
    z_\alpha ,\quad K_\alpha =\sqrt{\mbox{Re}z_\alpha +|z_\alpha |}. 
\end{equation}
By our definition $z$ is proportional to the complex $p^2$. In turn complex
$p^2$ reads $p^2=(p_0-i\eta '/2)^2-\mbox{\boldmath $p$}^2$. In the rest frame
$\mbox{\boldmath $p$}=0$ and then $p^2=(M-i\eta /2)^2$. Thus $\omega$ has a
meaning of mass and $\eta$ is related to the width of the resonance state. If 
$\eta$ is small one has $p^2=M^2-i\gamma$, where $\gamma =M\eta$. We used this
fact in paragraph 2. 

We plot resonance branches for $\rho$ and $a_1$ mesons in Figs.1-2 for two 
different values of quark masses: $m=390$\ MeV and $m=180$\ MeV, using the same
set of parameters ($\widehat{m}, m_\pi, \Lambda$) as in section~5 and varying 
$G_V$. Notice that in the case of the $\rho$ meson only the bound state branch 
has been known until now. At $p^2>4m^2$ this curve smoothly transforms into a 
``virtual state" section. This has been first observed in \cite{Takizawa:1991}.
In nonrelativistic scattering theory, the appearance of a ``virtual state" is 
directly related to the absence of a near-threshold resonance in the $s$-wave. 
In the relativistic theory, as we have found here by studying the NJL model, 
the resonance branch appears together with a bound state--``virtual state" 
branch as a separate curve. It is never smoothly connected with the bound state
branch and it has rather large values of the imaginary part $\eta_\rho$. For 
large values of quark mass $m\sim 400$\ MeV (see Fig.1) the $\rho$ meson 
solution belongs to the bound state branch. However, at relatively small 
values, $m\sim 200$\ MeV, as one can see from Fig.2, the $\rho$ meson solution 
is a resonance state in the quark--antiquark continuum. 

We plot in Figs.1-2 only the real parts, $\omega_\alpha$, of resonance 
solutions $m_\alpha =\omega_{\alpha}-i\eta_{\alpha}$. Therefore the 
intersection of the resonance branch with the ``virtual state" in Figs.1-2 is 
only a result of this projection on the real $m_\alpha$ axis. Really the 
crossing curves have different imaginary parts $\eta_\alpha$ in these points. 
We draw them in Figs.3-4. For $m=390$\ MeV we obtain for the empirical value 
of $f_\pi =93.3$\ MeV the $\rho$-meson bound state mass $m_\rho =776$\ MeV and
the complex mass for $a_1$, $m_{a_1}=1000-i263$\ MeV. 
For $m=180$\ MeV, the complex masses of resonance states which correspond to 
the empirical value of $f_\pi =93.3$\ MeV are $m_\rho =674-i292$\ MeV, and 
$m_{a_1}=802-i289$\ MeV. 
These values are a bit smaller than the experimental ones. To clarify the 
$f_\pi$ dependence of the result we indicate in Figs.1-2 boxes inside which the
$f_\pi$ value changes from $f_\pi =100$\ MeV on the left side to $f_\pi =92$\ 
MeV on their right side. The physical solutions lie inside this box. In the 
other figures we directly plot $f_\pi$ as a function of $G_V$. As discussed in 
section 4, we have also  checked our results via
the speed plot technique.
Very good agreement between the two methods was obtained.     

It is also interesting  to compare the results of the exact solutions of the
complex BS equation for the $\rho$ and the $a_1$ with the approximate 
estimations of section~5 (see Figs.3-4). We start the comparison for the case 
of a large quark mass $m=390$\ MeV in Fig.3. As one can see the expansion 
approach works quite well for the $a_1$ meson. The deviation in the results of 
the two approaches is smaller at higher values of $G_V$ where $\eta_{a_1}$ is 
smaller as it should be expected. We also plot here the $\eta_\rho$ values. It 
follows from section~3 that at $p^2<4m^2$, Eqs.(\ref{acs}) describe a bound 
state branch. At large values of the quark mass, we have $\mbox{Re}(z_{\rho}) 
\ll 1$, far away from the region of applicability of Eqs.(\ref{acs}) to 
resonance solutions. The resonance branch as an approximate solution cannot be 
found in this case. The physical branch for $\rho$ is a bound state solution, 
for which the exact solution fully coincides with the solution of 
Eqs.(\ref{acs}). In Figs. 4a and 4b we compare the full results with the
expansion approach at small quark mass values $m=180$\ MeV in the case 
of the $a_1$ and $\rho$ mesons respectively. In the case of the $\rho$-meson
the larger deviations observed between approximate and exact solutions at
$G_V>12\  \mbox{GeV}^{-2}$ values stem from the fact that
$\mbox{Re}(z_{\rho}) < 1$. In this region, $12 < G_V <20\  \mbox{GeV}^{-2}$,
the expansion curves interpolate between the resonance solutions and
the bound state branch (if it exists). In Fig.4b we do not plot the bound
state branch which starts at $G_V=20\  \mbox{GeV}^{-2}$. 
As mentioned above, Eqs.(\ref{acs}) are not appropriate to describe resonance
solutions in this region. The exact scalar resonances coincide with the ones 
obtained with the approximate ones (section~5) and are practically constant as 
functions of $G_V$.  

\vspace{1truecm}

\noindent {\bf 7.} As in ref.\cite{Takizawa:1991} we have studied here the 
resonance solutions in the extended NJL model in the quark--antiquark 
continuum. There is only one difference in our Lagrangian in comparison with 
that paper. We consider the model without the 't Hooft determinantel 
interaction. This interaction is important for solving the $U(1)$ anomaly 
problem and, therefore, essential for the calculation of $\eta -\eta '$ meson 
masses \cite{Bernard:1988}. Since we restrict our consideration to the case of 
$\rho$, $a_1$ and $\sigma$ mesons one can neglect these six quark vertices. In 
any case, the results found in our study are independent of the precise form of
the underlying NJL--type model. In ref.\cite{Takizawa:1991}, the authors
could not find the resonance solution in the case of the vector $\rho$ meson. 
Instead, the ``virtual state" has been found. In this letter we have studied 
this question more thoroughly. We have shown that the relativistic BS equation 
for the $\rho$ meson state possesses also a resonance solution. This solution 
has a rather large imaginary part and corresponds to the physical $\rho$ at 
small values of constituent quarks masses $m\sim 200$ MeV. The imaginary part 
is a consequence of the deconfinement property of the NJL model and should be 
considered an artefact of the model. Nevertheless, the existence of resonance 
solutions is a very important feature of the model. It means in particular that
the $\rho$ meson state in the model exists as an asymptotic field and 
contributes at large time even at small values of $m$, although the probability
for the detection of the particle decreases exponentially with a characteristic
decay time $\tau (\mbox{\boldmath $p$})=\omega (\mbox{\boldmath $p$})/\gamma$. 
The well-known Lagrangian (bosonization) approaches to the NJL model \cite{
Bijnens:1994} use this fact without proof. An asymptotic $\rho$ meson field is 
usually introduced in the bosonized NJL Lagrangian by the replacements of 
variables in the generating functional and finally, after the fixing of 
parameters, it happens often that the quark mass value is of the order of 
$200\div 300$\ MeV. This procedure has a justification only if this state 
really exists in the model as a resonance in a quark--antiquark continuum. Our 
work proves for the first time this important fact.

\vskip 0.4truecm
{\bf Acknowledgments}
One of us (V.B.) would like to thank V. Branchina and J. Alexandre for useful
discussions. It is a pleasure for A.B., B.H. and A.O. to thank E. van Beveren 
for stimulating discussions.

\baselineskip 12pt plus 2pt minus 2pt


\vspace{2cm}

\section*{Figure captions}
\vspace{0.5cm}

\begin{itemize}

\item[Fig.1] Dependence of the real part $\omega_\alpha ,\quad (\alpha =\rho$, 
and $a_1$) of the exact vector, $\rho$, and axial-vector, $a_1$, resonance 
solutions on $G_V$ at large values of constituent quark masses $m=390$\ MeV. 
The solid line corresponds to the bound state branch. It smoothly touches the 
$q\bar{q}$ threshold line, $2m$. The dashed lines indicate resonance branches 
of $\rho$ and $a_1$ mesons. The dotted line is a ``virtual-state" solution. The
physical $\rho$ is a bound state solution. The solid box indicates the region 
bounded by the values of a pion decay constant between $f_\pi =100$\ MeV (left 
side of the box) and $f_\pi =92$\ MeV (right side) where the physical $\rho$ 
meson lies.
\vspace{0.5cm}

\item[Fig.2] Same as Fig.1 for the small constituent quark mass $m=180$\ MeV. 

\vspace{0.5cm}

\item[Fig.3] Comparison of the exact solutions (dashed lines) of the $a_1$ 
meson BS equation for the quark mass $m=390$\ MeV at different values of $G_V$ 
with the corresponding solutions of the Eqs.(\ref{acs}) (dotted lines). 
Indicated is also the exact imaginary part $\eta_\rho$ and $f_\pi$.
For more details see text.

\vspace{0.5cm}

\item[Fig.4] (a) Comparison of the exact solutions of the  $a_1$ meson BS 
equation for the quark mass $m=180$\ MeV as functions of of $G_V$ with the 
corresponding solutions of the Eqs.(\ref{acs}), same notation as in Fig.3. 
(b) Same as (a) for the $\rho$ meson. Indicated is also $f_\pi$. 

\end{itemize}

\newpage
$\;$

\vskip 3cm

\begin{figure}[bht]
\centerline{
\epsffile{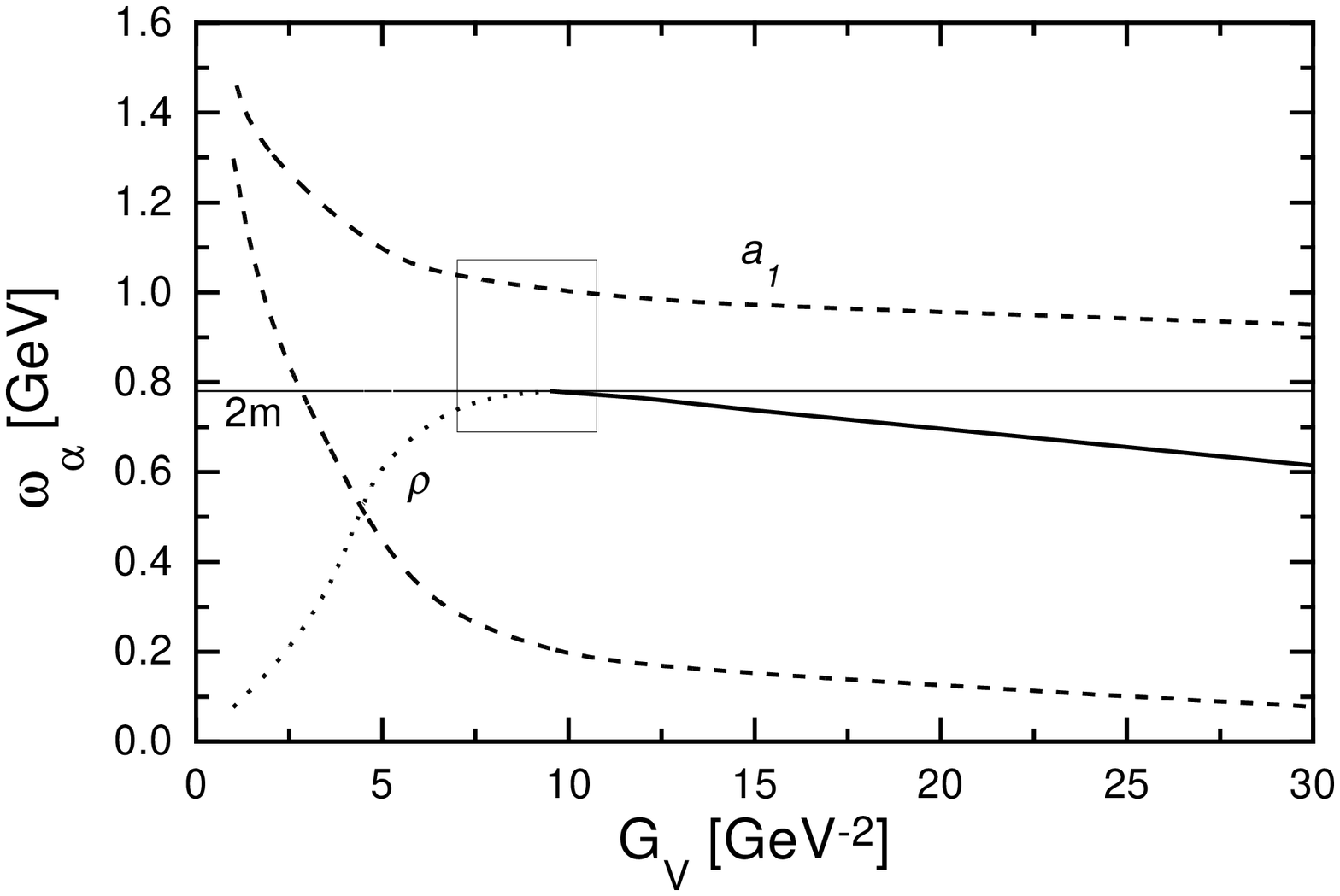}
}
\vskip 2.5cm

\centerline{\Large Figure 1}
\end{figure}
 

\newpage

$\,$

\vskip 2cm

\begin{figure}[bht]
\centerline{
%
%
\epsfysize=5in
\epsffile{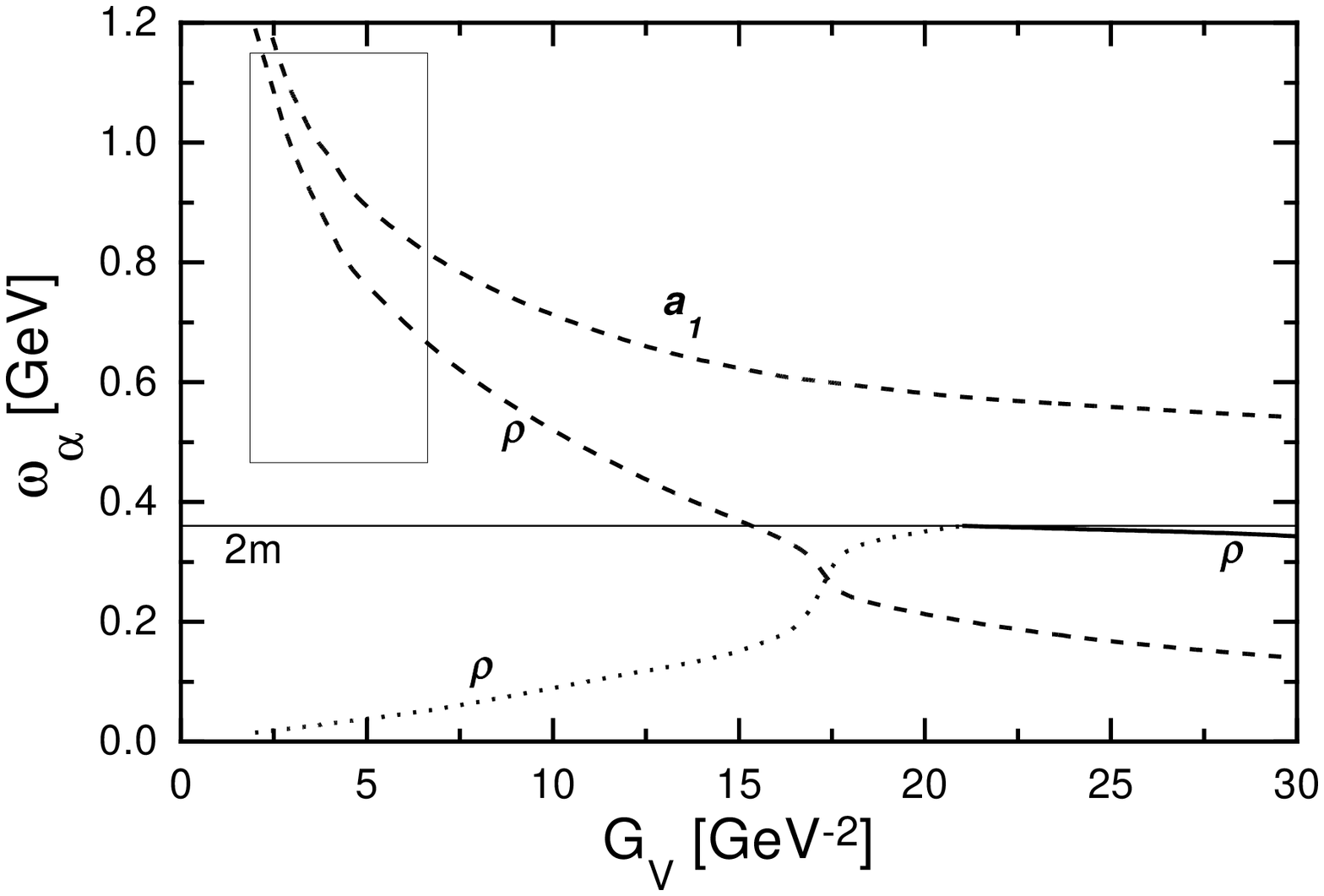}
}
\vskip 2.5cm

\centerline{\Large Figure 2}
\end{figure}
 

\newpage
$\;$

\vskip 3cm

\begin{figure}[bht]
\centerline{
\epsffile{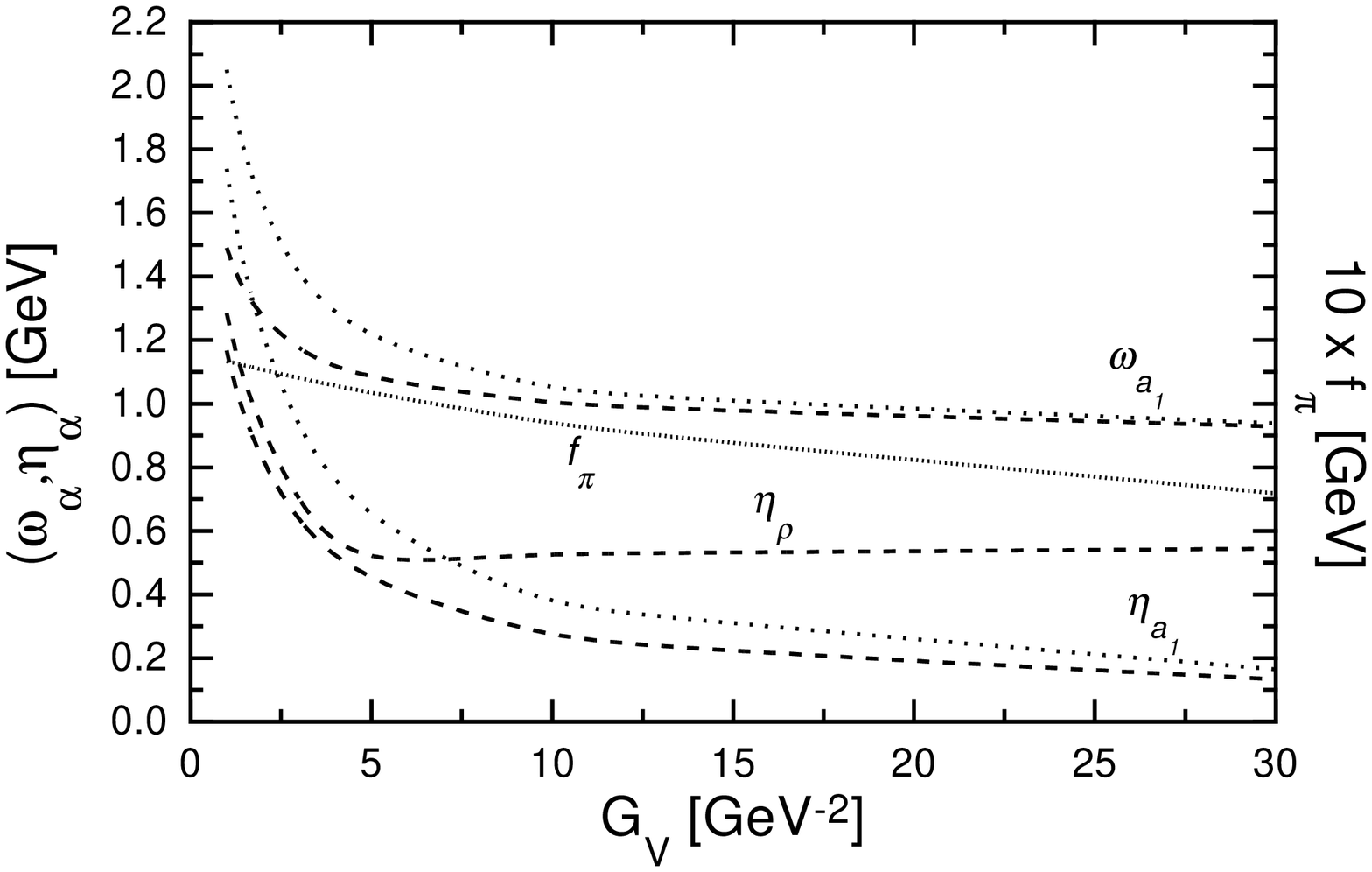}
}
\vskip 2.5cm

\centerline{\Large Figure 3}
\end{figure}
 

\newpage

$\,$

\vskip 2cm

\begin{figure}[bht]
\centerline{
%
%
\epsfysize=5in
\epsffile{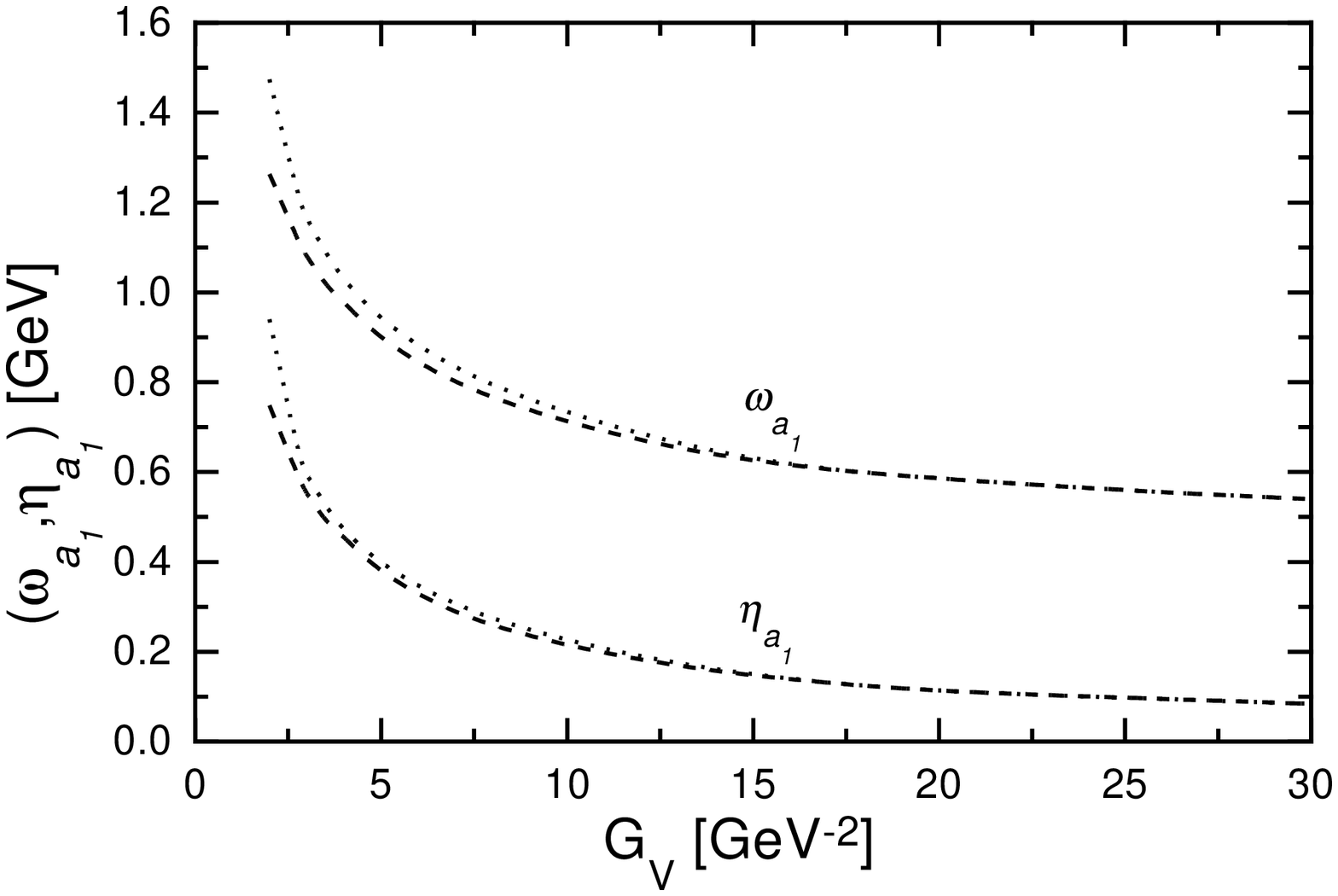}
}
\vskip 2.5cm

\centerline{\Large Figure 4a}
\end{figure}
 

\newpage

$\,$

\vskip 2cm

\begin{figure}[bht]
\centerline{
%
%
\epsfysize=5in
\epsffile{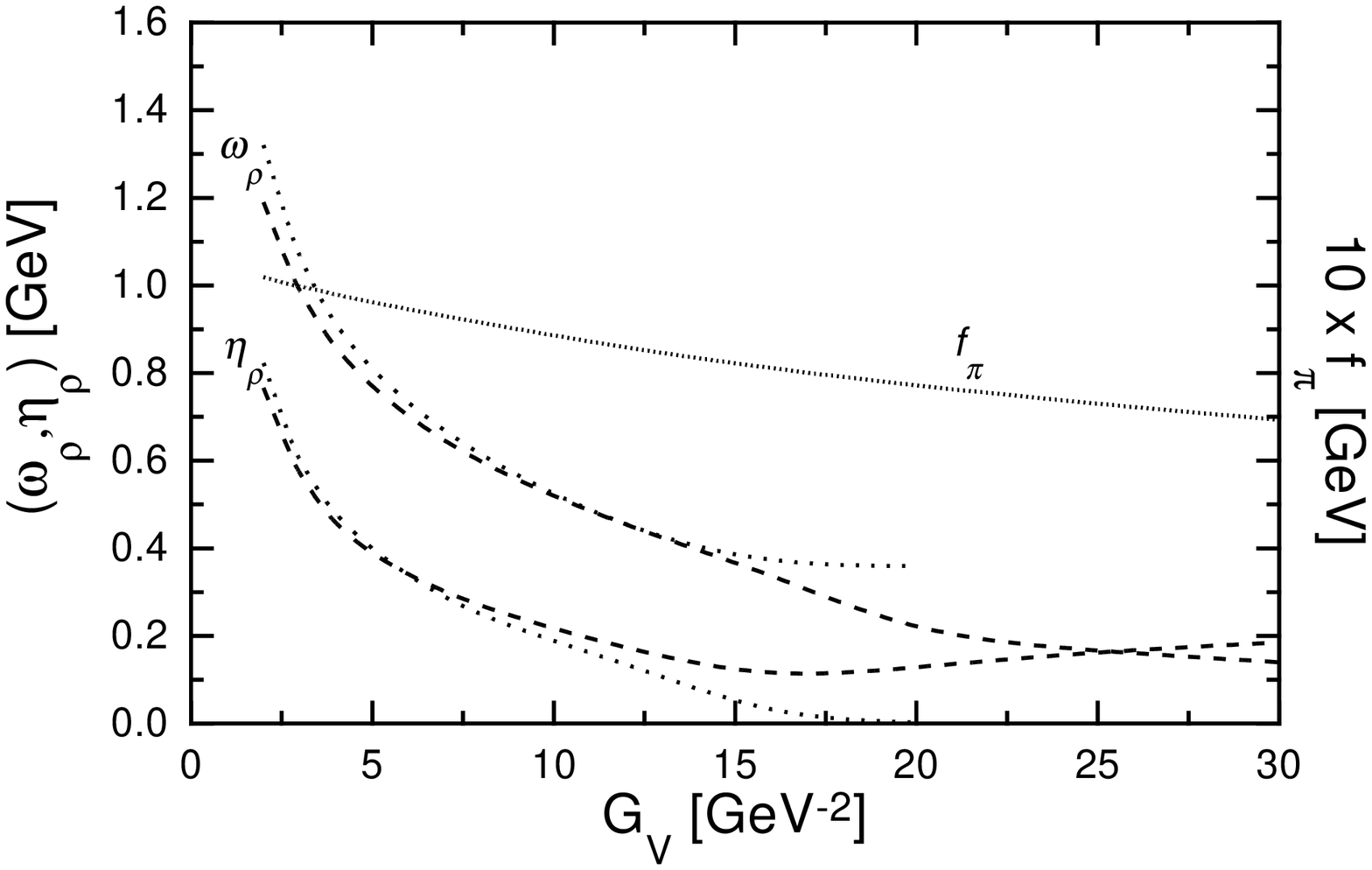}
}
\vskip 2.5cm

\centerline{\Large Figure 4b}
\end{figure}
 

\end{document}